\documentclass[reprint,nofootinbib,aps,superscriptaddress,amsmath,amssymb,onecolumn,11pt]{revtex4-2}

\usepackage{graphicx}
\usepackage{dcolumn}
\usepackage{bm}
\usepackage{braket}
\usepackage{breqn}
\usepackage{mathrsfs}
\usepackage{color}

\usepackage[colorlinks=true,linkcolor=red,citecolor=blue]{hyperref}

\begin{document}

\title{Charged superradiant instability in a spherical regular black hole}

\author{Yizhi Zhan}
\email{niannian$\_$12138@163.com}
\author{Hengyu Xu}
\email{xuhengyu0501@outlook.com}
\author{Shao-Jun Zhang}
\email{sjzhang@zjut.edu.cn (corresponding author)}
\affiliation{Institute for Theoretical Physics and Cosmology$,$ Zhejiang University of Technology$,$ Hangzhou 310032$,$ China}
\affiliation{School of Physics$,$ Zhejiang University of Technology$,$ Hangzhou 310032$,$ China}
\date{\today}

\begin{abstract}
 We examine the stability of a spherically symmetric regular black hole when subjected to perturbations from a charged scalar field. This particular black hole is constructed by deforming the Minkowski spacetime. It has been observed that the charged superradiant instability arises only within a specific range of the deformation parameter, potentially resulting in an instability growth rate with a maximum magnitude of approximately $\text{Im} (M \omega) \sim 10^{-3}$. This growth rate significantly exceeds the instability identified in Ay\'{o}n-Beato-Garc\'{i}a (ABG) black holes discussed in prior research, suggesting a notable timescale for detecting this phenomenon in astrophysical scenarios. Additionally, we conduct a thorough investigation into how the three parameters of the model influence the onset and intensity of the instability. Our analysis offers further insights into the possible emergence of this instability in spherical regular black holes and its association with the nonlinear effects of the electromagnetic field. 
    
\end{abstract}

\maketitle
  
\section{Introduction}

Recent years have seen significant advancements in astronomical observation, with a notable breakthrough being the groundbreaking detection of gravitational waves \cite{LIGOScientific:2016aoc,LIGOScientific:2016sjg} and the remarkable imaging of the black hole shadow \cite{EventHorizonTelescope:2019dse,EventHorizonTelescope:2019ggy}. These achievements have propelled black holes to the forefront of current scientific research. Predicted by Einstein's theory of gravity, black holes represent the most compact objects in the universe and are the eventual result of the collapse of extremely massive stars \cite{Tolman:1939jz,Oppenheimer:1939ue,Joshi:2011rlc,Christodoulou:1984mz,Penrose:1969pc,Joshi:2008zz}. They are anticipated to have a pivotal role in the development of galactic structures and the evolution of the universe.

The singularity problem stands out as a key aspect among the distinct features of black holes, playing a crucial role in elucidating black hole physics. Whether examined within the confines of Einstein's gravity theory or in alternative gravitational frameworks, the majority of established black hole solutions are characterized by the existence of singularities. Notably, the singularity theorems put forth by Penrose, Hawking, and others assert that, in classical gravity scenarios, singularities represent the unavoidable outcome of matter succumbing to strong gravitational forces during collapse \cite{Penrose:1964wq,Hawking:1973uf,Senovilla:1998oua}. Consequently, delving into the intricacies of black hole singularities is essential for comprehending the definitive outcome of gravitational collapse and the intricate process of black hole genesis.

Singularities are commonly considered non-physical within the classical theoretical framework due to their infinite energy density, prompting researchers to seek resolution through the inclusion of quantum effects in matter or gravity. While singularity theorems rely on strict conditions like the null energy condition for matter, allowing for the possibility of singularity-free black holes. To address the issue of non-physical singularities, a significant effort is being made by researchers to develop regular black hole solutions. This endeavor dates back to 1968 when Bardeen introduced a regular black hole model that adhered to the weak energy condition \cite{1968qtr..conf...87B}, circumventing the singularity theorems. Subsequently, Ay\'{o}n-Beato and Garc\'{i}a established a class of regular black holes known as ABG black holes \cite{Ayon-Beato:1998hmi, Ayon-Beato:1999kuh}, featuring spherical charged configurations within the framework of Einstein-nonlinear electromagnetic fields. Further advancements led them to realize the Bardeen black hole as an exact solution within some Einstein-nonlinear-Maxwell theory \cite{Ayon-Beato:2000mjt}. These breakthroughs have inspired the development of various regular black hole models predominantly supported by nonlinear electromagnetic fields (see for example \cite{Fan:2016hvf, Ghosh:2022gka, Lan:2023cvz, Torres:2022twv, Bronnikov:2022ofk, Bueno:2024dgm} and references therein), which serve as an avenue to incorporate quantum effects in the electromagnetic field theory \cite{Heisenberg:1936nmg,Born:1934gh,Polchinski:1998rq,Polchinski:1998rr}. Extensive investigations have been conducted on the dynamic behaviors, thermodynamic properties, and astrophysical implications of these black holes. The distinct observational effects exhibited by regular black holes compared to conventional ones have been a subject of interest, as indicated by recent studies and reviews on the topic \cite {Allahyari:2019jqz,Vagnozzi:2022moj,Lan:2023cvz,Torres:2022twv,Bronnikov:2022ofk,Cai:2020kue,Cai:2021ele,Yang:2022uze,Meng:2022oxg,Pedrotti:2024znu}.

The authors of Refs. \cite{Hod:2024aen,dePaula:2024xnd,Dolan:2024qqr} recently conducted an extensive examination of the superradiance phenomenon in ABG black holes subjected to charged scalar field perturbations. Their findings suggest that within certain parameter ranges, ABG black holes display superradiant instability, a characteristic distinct from the well-established behavior of Reissner-Nordstr\"om (RN) black holes. The study of superradiance and its associated instability in black holes has been a focal point of research in black hole physics for a significant period. Superradiant instability typically manifests in rotating black holes \cite{Damour:1976kh,Detweiler:1980uk,Zouros:1979iw,Cardoso:2005vk, Dolan:2007mj, Konoplya:2013rxa}. However, for non-rotating, spherically symmetric RN black holes, superradiance can occur under charged scalar field perturbations but leading no instability \cite{Hod:2012wmy,Hod:2013nn}, unless the spacetime has a special boundary or an artificial boundary condition is imposed \cite{Herdeiro:2013pia, Zhu:2014sya, Konoplya:2014lha, Dolan:2015dha, Dias:2018zjg, Davey:2021oye, Richarte:2021fbi, Feiteira:2024awb}. The research outcomes in the Refs. \cite{Hod:2024aen,dePaula:2024xnd,Dolan:2024qqr} introduce a novel scenario: when accounting for the nonlinear effects of the electromagnetic field, spherically symmetric charged regular black holes may also exhibit superradiant instability. The superradiant instability of black holes holds significant astronomical observational implications. This mechanism results in the accumulation of scalar particles around the black hole, giving rise to observable effects such as alterations in gravitational waveforms and interference with the imaging of the black hole shadow \cite{Arvanitaki:2009fg,Arvanitaki:2010sy,Brito:2014wla,Cunha:2015yba,Vincent:2016sjq,Cunha:2019ikd,Creci:2020mfg}. These observable effects offer valuable insights and avenues for exploring exotic particles beyond the standard model and furthering our comprehension of the physical characteristics of black holes. See more discussions on this phenomenon and its astrophysical implications, please refer to the review \cite{Brito:2015oca, Konoplya:2011qq} and references therein. Thus, the research findings in Refs. \cite{Hod:2024aen,dePaula:2024xnd,Dolan:2024qqr} not only unveil the superradiant instability of ABG black holes under specific circumstances but also present a potential observational approach for identifying and investigating the properties of regular black holes.

The inquiry arises as to whether the phenomenon of superradiant instability is exclusive to ABG black holes or is a common occurrence among regular black holes. This study delves into this question by exploring a distinct class of regular black holes \cite{Ovalle:2023ref} generated through the ``gravitational-decoupling" (GD) technique \cite{Ovalle:2017fgl, Ovalle:2018gic}, possessing unique characteristics. The spacetime of these black holes serves as an interpolation between Minkowski and Schwarzschild spacetime, characterized by a deformation parameter and potentially supported by a nonlinear electromagnetic field. Through a detailed analysis of the behavior of charged scalar field perturbations around such black holes, it is observed that the presence and strength of superradiant instability are closely linked to the deformation parameter. Remarkably, when the deformation parameter falls within a specific range, the intensity of superradiant instability can escalate significantly, surpassing that of ABG black holes. These findings provide evidence that regular black holes typically display superradiant instability.

This paper is structured as follows: Section II provides a concise overview of the regular black hole under consideration. Section III delves into the equation of motion governing the tested charged scalar field. In Section IV, an examination of the parameter space is conducted to determine the permissible region for initiating the superradiant instability. Section V employs a numerical approach to identify the superradiant unstable modes. The concluding section encompasses a summary and discussions. The study employs natural units with $G = c = \hbar = 4 \pi \epsilon_0 = 1$ throughout, where $G, c, \hbar$ and $\epsilon_0$ are the Newtonian gravitational constant, the photon velocity in vacuum, the induced Planck constant and the vacuum permittivity, respectively.

\section{The spacetime}

We consider the static and spherically symmetric regular black hole proposed in Ref. \cite{Ovalle:2023ref}. The line element is
   \begin{align}
    ds^2 &=-f(r)dt^2+f(r)^{-1}dr^2+r^2d\Omega^2,\\
    f(r) &=1-\frac{2M}{r}+\frac{e^{-\alpha r/M}}{rM} \left(\alpha^2r^2+2M\alpha r+2M^2 \right),
\end{align}
where $M$ is the ADM mass of the black hole and $\alpha$ is a deformation parameter. It is constructed through the deformation of the Schwarzschild spacetime with the GD method \cite{Ovalle:2017fgl,Ovalle:2018gic}. The metric represents a spacetime interpolating the Minkowski spacetime and the Schwarzschild black hole: In the limit of $\alpha \rightarrow 0$, it reduces to the Minkowski metric; while in the limit of $\alpha \rightarrow \infty$, it reduces to the standard Schwarzschild metric. It should also be noted that the deformation of this spacetime with respect to the Schwarzschild metric takes an exponential decaying form, so that the deformation mainly affects the small-$r$ region but leaves the large-$r$ region little affected. The spacetime has a de-Sitter core and is regular at $r=0$ which can be seen by calculating the typical curvature invariants, for example,
\begin{align}
    R &= \frac{\alpha^3 (4M-\alpha r)}{M^3} e^{-\alpha r/M},\\
    R_{\mu\nu} R^{\mu\nu} &= \frac{\alpha^6 (8M^2-4 \alpha r M+\alpha^2 r^2)}{2 M^6} e^{-2\alpha r/M},\\
    R_{\mu\nu\rho\sigma} R^{\mu\nu\rho\sigma} &\simeq \frac{8\alpha^6}{3M^4}-\frac{20\alpha^7 r}{3M^5}+\frac{35\alpha^8 r^2}{4M^6},\qquad r\rightarrow 0.
\end{align}

As most regular black holes, this metric can be sourced by some kind of nonlinear electromagnetic field which takes a static configuration $A_\mu = (-\Phi (r), 0, 0, 0)$, where the electric potential is \cite{Ovalle:2023ref}
\begin{align}
    \Phi(r) = \frac{e^{-\alpha r /M}(6M^3+6M^2 r \alpha+3Mr^2 \alpha^2+r^3 \alpha^3)}{4M^2 \alpha}.
\end{align}
It asymptotically takes a Yukawa-like form rather than a Maxwell-like form, which is very different from that of ABG black holes \cite{Ayon-Beato:1998hmi,Ayon-Beato:1999kuh}. So the ``nonlinear electromagnetic field" may not be considered as the usual nonlinear extension of Maxwell's theory simply, rather it is more likely associated with some kind of short-range interaction. The action from which to derive this solution is rather involved and not relevant to our present study, so we will not display it here. For more details, one can refer to Ref. \cite{Ovalle:2023ref}.

\begin{figure}[!htbp]
     \includegraphics[width=0.45\textwidth]{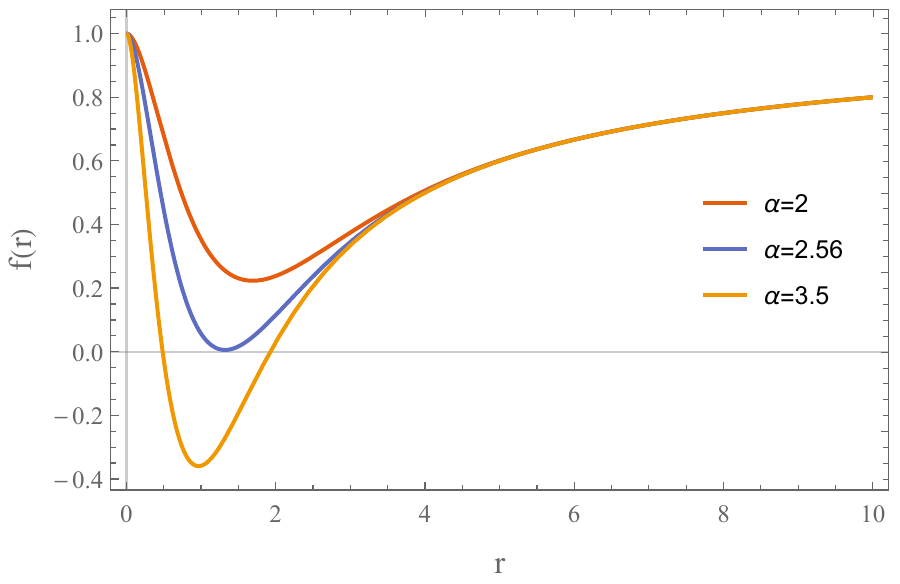}\quad
     \includegraphics[width=0.45\textwidth]{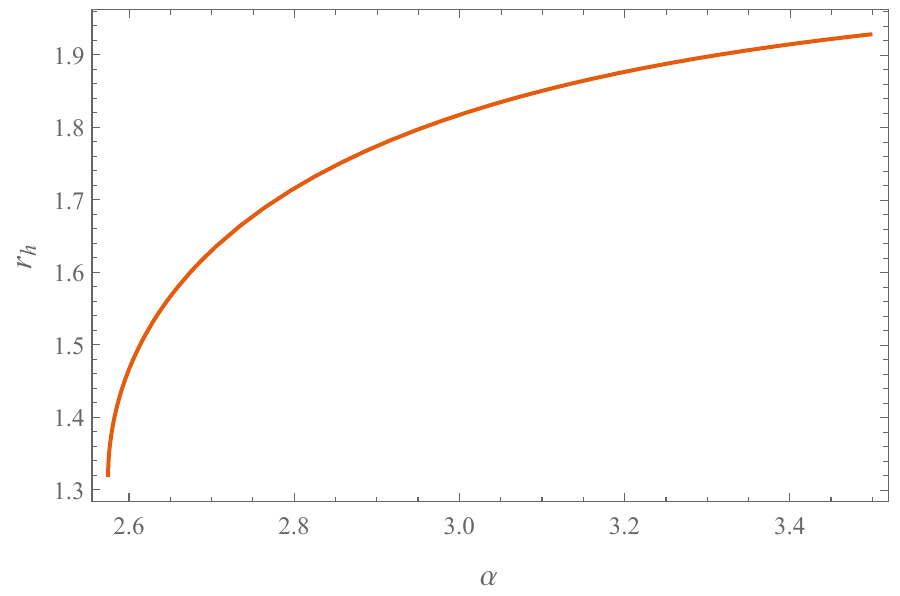}   
    \caption{{\em Left:} Profile of the metric function $f(r)$ for different values of $\alpha$. {\em Right:} Horizon radius $r_h$ as a function of $\alpha$. The black hole mass is set to be $M=1$, so all physical quantities are measured in units of $M$.}   \label{fig:Horizon}
\end{figure}

The event horizon $r=r_h$ of the black hole is given by the largest root of $f(r)=0$, which depends on the value of $\alpha$. In Fig. \ref{fig:Horizon}, we plot the profile of the metric function $f(r)$ and the horizon radius $r_h$ for various $\alpha$. From the left panel, one can see that there exists a threshold value of $\alpha = \alpha_c \simeq 2.56$ only above which $f(r)$ has zero points. This means that only when $\alpha \geq \alpha_c$, the metric represents a black hole and has two horizons when $\alpha > \alpha_c$. Moreover, as $r\rightarrow 0$ or $r \rightarrow \infty$, $f(r) \rightarrow 1$. From the right panel, one can see that the horizon radius $r_h$ increases with the increase of $\alpha$, and approaches the Schwarzschild radius $r_h = 2 M$ in the limit of $\alpha \rightarrow \infty$.

\section{Test charged scalar field}

Now we consider a test charged scalar field $\Psi$ propagating in spacetime and search for the possible unstable modes under superradiance. Assume that the scalar field is minimally coupled to the electromagnetic field, so its dynamic is governed by the Klein-Gordon equation, 
   \begin{align}
    (\nabla_\nu-iqA_\nu)(\nabla^\nu-iqA^\nu)\Psi-\mu^2 \Psi=0,
\end{align}
where $\mu$ and $q$ represent the mass and charge parameters of the field, respectively. Due to the spherical symmetry of the spacetime, the scalar field can be decomposed as
   \begin{align}
    \Psi=\frac{1}{r}u_{\omega \ell}(r)Y_{\ell m}(\theta,\phi)e^{-i\omega t},
\end{align}
where $Y_{\ell m} (\theta, \phi)$ are the spherical harmonics. For simplicity, we will henceforth omit the subscripts from $u_{\omega \ell}(r)$. Substituting the above decomposition into the field equation, one derives a radial equation of the form 
   \begin{align}
    \left\{f(r)\frac{d}{dr}\left[f(r)\frac{d}{dr}\right]-V(r)\right\}u(r)=0, \label{RadialEq1}
\end{align}
where the effective potential $V(r)$ is given by 
   \begin{align}
    V(r)\equiv f(r)\left[\mu^2+\frac{1}{r} \frac{df(r)}{dr}+\frac{\ell(\ell+1)}{r^2}\right]-\left[\omega-q\Phi(r)\right]^2. \label{EffectivePotential}
\end{align}
It is easy to see that the effective potential has the following asymptotic behaviors,
\begin{eqnarray}
V(r) \rightarrow \left\{
\begin{array}{cc}
- \tilde{\omega}^2, & \quad r \rightarrow r_h,\\
k^2, & \quad r \rightarrow \infty,
\end{array}
\right.
\end{eqnarray}
where $\tilde{\omega}\equiv\omega-\omega_c, \ \omega_c\equiv q\Phi(r_h)$, and $k \equiv \sqrt{\mu^2 - \omega^2}$. 

To solve the radial equation (\ref{RadialEq1}) to obtain the frequency $\omega$ and the associated modes $u(r)$, we need to impose physical boundary conditions: the wave should be purely ingoing near the horizon, while bounded at infinity to trigger superradiant instability. With the tortoise coordinate defined as $d r_\ast=dr/f(r)$, the radial equation (\ref{RadialEq1}) can be cast into a Schr\"{o}dinger-like form
\begin{align}
    \left\{\frac{d^2}{dr_\ast^2} -V(r)\right\}u(r)=0. \label{RadialEq2}
\end{align}
From it, one can arrive at the following boundary conditions,
\begin{eqnarray}
u(r) \rightarrow \left\{
\begin{array}{cc}
e^{- i \tilde{\omega} r_\ast}, & \quad r \rightarrow r_h,\\
e^{- k r_\ast}, & \quad r \rightarrow \infty,
\end{array}
\right.\label{BoundaryConditions}
\end{eqnarray}

\section{Analysis on parameter space}
\subsection{The hydrogenic approximation}

Let us first solve the radial equation (\ref{RadialEq1}) in the hydrogenic approximation with $M \mu \ll 1$. This approximation can give us a good estimate of the real part of $\omega$. By expanding the radial equation at large $r$, we have 
\begin{align}
    \left[\frac{d^2}{dr^2} - k^2 + \frac{2 M \mu^2}{r} - \frac{\ell (\ell +1)}{r^2} \right] u(r) =0.
\end{align}
Solving this equation with bounded boundary condition at infinity, one has
\begin{align}
    u(r) = e^{-k r} (k r)^{\ell +1} U \left(\nu, 2\ell+2, 2 k r\right),
\end{align}
where $U \left(\nu, 2\ell+2, 2 k r \right)$ is the confluent hypergeometric function and $\nu \equiv \ell +1 - \frac{M \mu^2}{k}$. Bound states demand that $\nu$ takes non-positive integer values, so we have 
\begin{align}
    \frac{M \mu^2}{k} = \ell + 1 + n \equiv \bar{n},
\end{align}
where $n=0, 1, 2, \cdots$ and $\bar{n} = 1, 2, \cdots$. From it, we get the hydrogenic spectrum
\begin{align}
    \omega = \mu \sqrt{1- \frac{M^2 \mu^2}{\bar{n}^2}} \approx \mu \left( 1-  \frac{M^2 \mu^2}{2 \bar{n}^2}\right). \label{HydrogenicSpectrum}
\end{align}

\subsection{Analysis on parameter space}

Before diving into numerical calculations to search for superradiant unstable modes, let us first do a primary analysis in parameter space to estimate the allowed region to trigger the instability. There are three free parameters in our model, $\{\alpha, \mu, q\}$, which determine the modes. 

To trigger superradiant instability, two necessary conditions should be satisfied simultaneously: (i) There should be superradiance, which demands the frequency of the wave to be lower than the critical value $\omega_c$, that is  \cite{Bekenstein:1973mi}
\begin{equation}
    0 < \omega < \omega_c.
\end{equation}
(ii) The effective potential $V(r)$ should have a potential well to trap the superradiant modes to form bound states, and thus superradiance is repeated to trigger instability.

\begin{figure}[!htbp]
\centering
\includegraphics[width=0.7\textwidth]{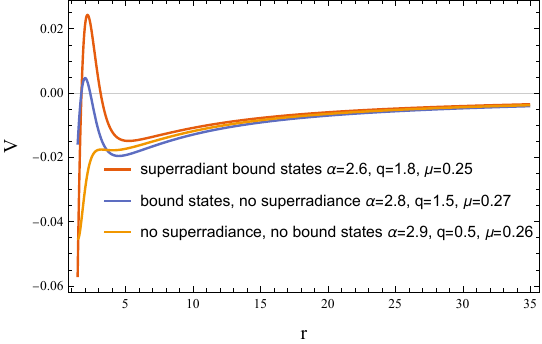}
\caption{Typical profile of the effective potential $V(r)$. We set $\ell=n=0$. The frequency $\omega$ is estimated with the hydrogenic approximation (\ref{HydrogenicSpectrum}). To trap the wave to form a bound state, there should be a potential well as well as a high-enough potential barrier.}
\label{EffectivePotential}
\end{figure}

Let us analyze the second condition in more detail. Compared with the standard Schrodinger equation, intuitively we can interpret the physical picture of the radial equation (\ref{RadialEq2}) as a particle with zero total energy moving in the potential $V(r)$. The effective potential $V(r)$ takes a typical profile as shown in  Fig. \ref{EffectivePotential}, from it one can see that it may possess a potential well as well as a potential barrier near the horizon. To trap the particle in the potential well to form a bound state, $V(r)$ should satisfy the following two conditions simultaneously: (i) $V(r) >0$ at infinity to keep the particle from escaping to infinity, which demands $\omega < \mu$; (ii) The potential barrier near the horizon should be higher than zero to keep the particle from falling into the black hole. 

\begin{figure}[!htbp]
\centering
\includegraphics[width=0.46\textwidth]{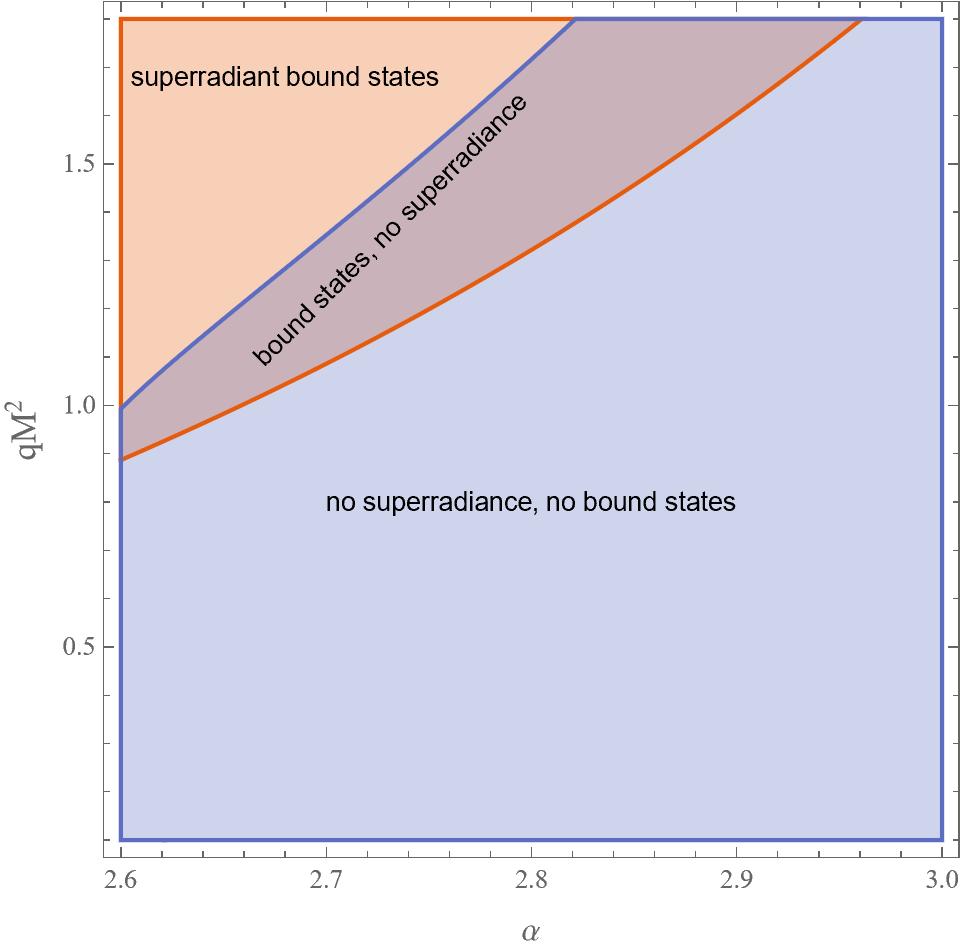}

\vspace{0.5cm}

\includegraphics[width=0.46\textwidth]{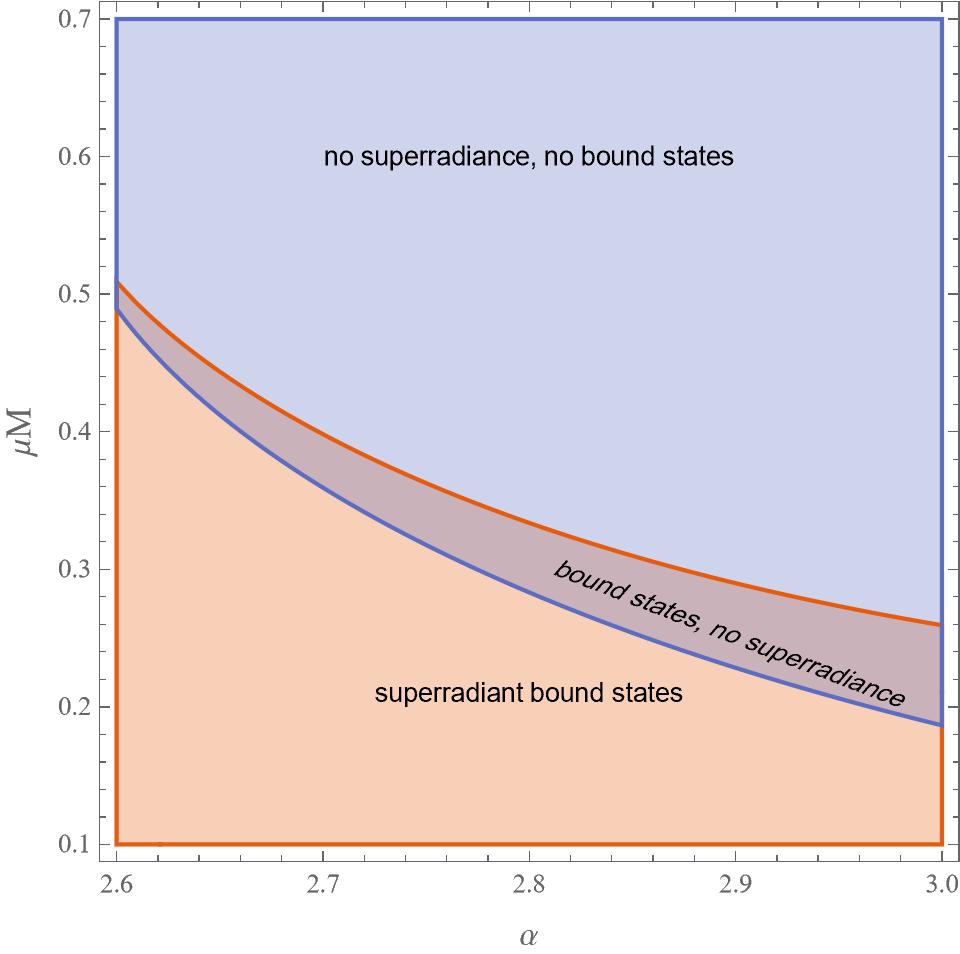}\quad
\includegraphics[width=0.46\textwidth]{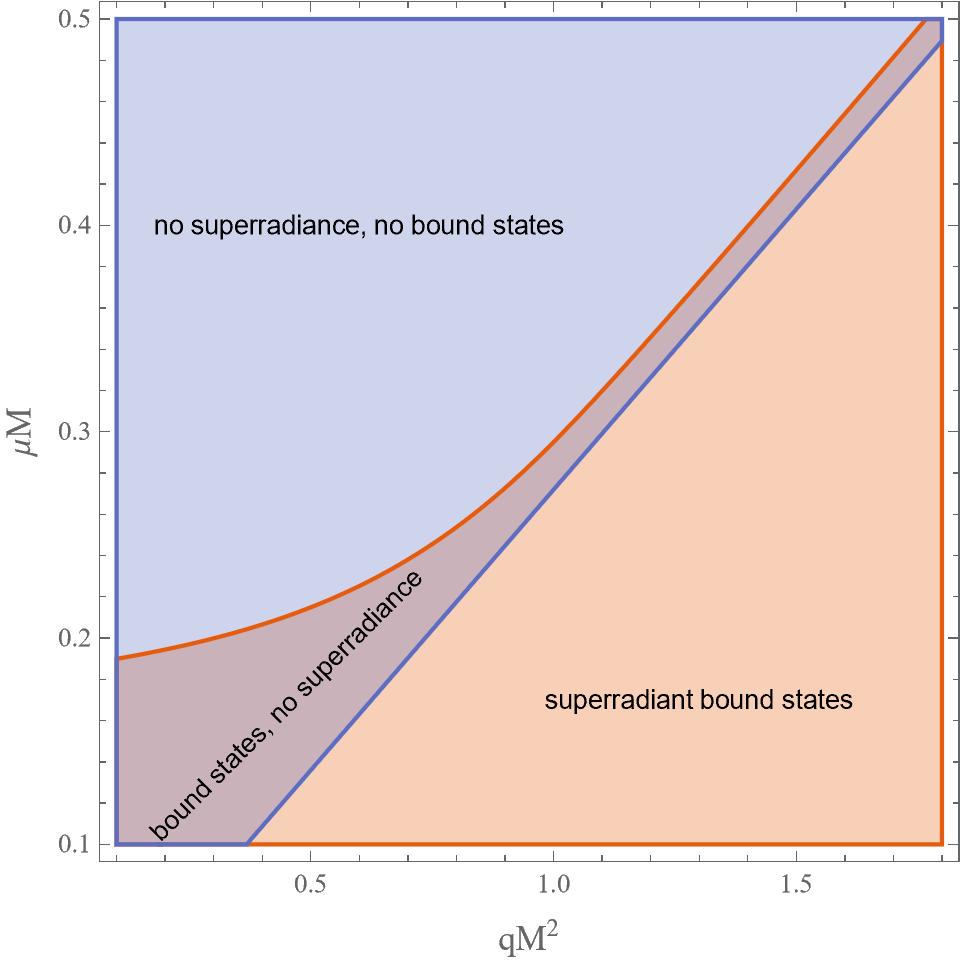}
\caption{Allowed regions for the possible existence of superradiant bound states in parameter spaces consisting of two out of the three parameters, $\{\alpha, \mu, q\}$. In the three panels, the remaining parameter is fixed as $\mu = 0.27$ (top), $q=1.8$ (bottom-left) and $\alpha=2.6$ (bottom-right), respectively. We set $\ell=n=0$. The frequency $\omega$ is estimated with the hydrogenic approximation (\ref{HydrogenicSpectrum}).}
\label{AllowedRegion}
\end{figure}

With the above necessary conditions and the help of the hydrogenic approximation (\ref{HydrogenicSpectrum}), one can give an estimation of the possible allowed region in parameter spaces for triggering superradiant instability. Typical samples are shown in Fig. \ref{AllowedRegion} with $\ell = n =0$, from which one can read off the qualitative influences of the interplay of the three parameters, $\{\alpha, \mu, q\}$, on the onset of the superradiant instability. Superradiant instability is observed in a narrow interval of $\alpha$ values, specifically within the range of $[\alpha_c, \alpha_0]$, when $q$ or $\mu$ are fixed, as evidenced from the top and bottom-left panels. The upper limit $\alpha_0$ is positively correlated with $q$ but inversely related to $\mu$. Similarly, there exist lower limits $q_c$ and $\mu_c$ for triggering superradiant instability when $\alpha$ is fixed, with $q_c$ increasing and $\mu_c$ decreasing as $\alpha$ increases. In the bottom-right panel, an upper limit $\mu_c$ on $\mu$ for inducing superradiant instability is observed when $q$ is fixed, with $\mu_c$ increasing as $q$ increases. Conversely, for a fixed $\mu$, there exists a lower limit $q_c$ on $q$ to trigger superradiant instability, with $q_c$ increasing as $\mu$ increases.

\section{Numerical Results}
\subsection{Numerical strategy}

Now we apply the numerical method of direct integration \cite{Pani:2013pma} to solve the radial equation (\ref{RadialEq1}) to calculate precise $\omega$. The details of the method are as follows. With the boundary conditions (\ref{BoundaryConditions}), the radial function $u(r)$ can be expanded in series form both near the horizon and the infinity respectively,
\begin{align}
    u(r) &=(r-r_h)^\sigma \sum_{j=0}^{N} u_{j} (r-r_h)^j, \qquad r\rightarrow r_h,\label{HorizonExpansion}\\
    u(r) &=r^c e^{- k r}\sum_{j=0}^{N} \frac{\tilde{u}_{j}}{r^j}, \qquad r\rightarrow \infty,\label{InfinityExpansion}
\end{align}
where $\sigma =- i \tilde{\omega} /f'(r_h)$ and $c = M (\omega^2 -k^2) /k$. $N$ is the truncation order. The expansion series coefficients $u_j$ and $\tilde{u}_j$ are to be determined. The metric function $f(r)$ and electric potential $\Phi(r)$ can also be expanded in a similar form. Inserting these expansions into the radial equation (\ref{RadialEq1}), the higher-order coefficients $u_j$ ($\tilde{u}_j$) with $j\geq1$ can be solved in terms of $u_0$ ($\tilde{u}_0$).  Furthermore, considering that the radial equation is linear in $u(r)$, we can fix $u_0=1$. So at last there remains only one free coefficient $\tilde{u}_0$.

In practical numerical calculations, the infinite radial domain should be truncated. We take $r \in [r_{\rm hn}, r_{\rm INF}]$, where $r_{\rm hn}=r_h (1 + \epsilon)$ is some point close to the horizon with $\epsilon$ being a small parameter, and $r_{\rm INF}$ is some large cutoff. Using the two expansions (\ref{HorizonExpansion}) (\ref{InfinityExpansion}) as boundary data at the two ends respectively, the radial equation (\ref{RadialEq1}) can be integrated either from $r_{\rm hn}$ or from $r_{\rm INF}$. We use a matching scheme to solve the equation. Namely, we integrate the equation from the two ends simultaneously and match the two solutions at some intermediate point $r=r_m$. Given the values of the parameters in the model, $\{\alpha, \mu, q\}$, the two solutions depend on $\{\omega, \tilde{u}_0\}$. Requiring smooth matching of the two solutions at $r_m$ will yield the desired frequency $\omega$ (as well as $\tilde{u}_0$). Typically, we take $N=3, \epsilon=10^{-3}$ and $r_{\rm INF}=1000 r_h$. We focus on the $\ell=0$ modes, which exhibit the strongest instability.

\subsection{Results}

By using the numerical method mentioned above, superradiant unstable modes have been found in certain parameter regions. In this section, we show the influences of the three parameters, $\{\alpha, q, \mu\}$, on the unstable modes. 

\begin{figure}[!htbp]
\centering
\includegraphics[width=0.64\textwidth]{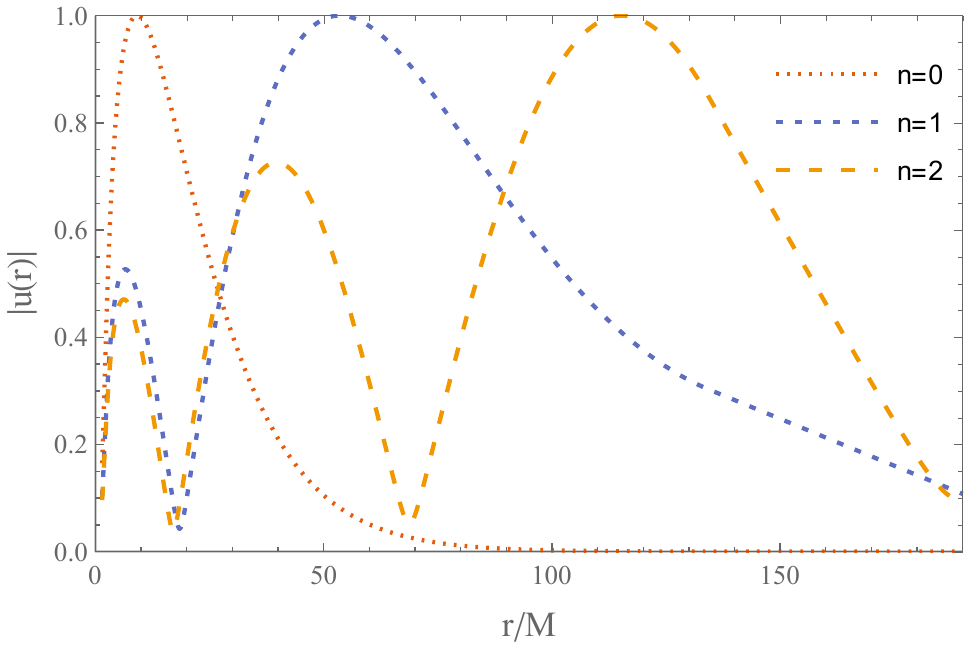}
\caption{Typical radial profiles of superradiant bound states for the first three modes $n=0, 1$ and $2$. Here $q/M = 1.8
, M\mu=0.27$ and $\ell=0$.}
\label{u(r)}
\end{figure}

Typical radial profiles $u(r)$ of the superradiant bound states are shown in Fig. \ref{u(r)} for the first three tones, $n=0, 1$ and $2$. Note that $n$ is the number of nodes $u(r)$ possesses in the radial direction. From the figure, one can see that the peak point of $|u(r)|$ shifts to larger $r$ as $n$ increases. This is expected and is consistent with the hydrogenic approximation, as in a hydrogen atom higher energy levels correspond to larger classical orbital radii. 

\begin{figure}[!htbp]
\centering
\includegraphics[width=0.46\textwidth]{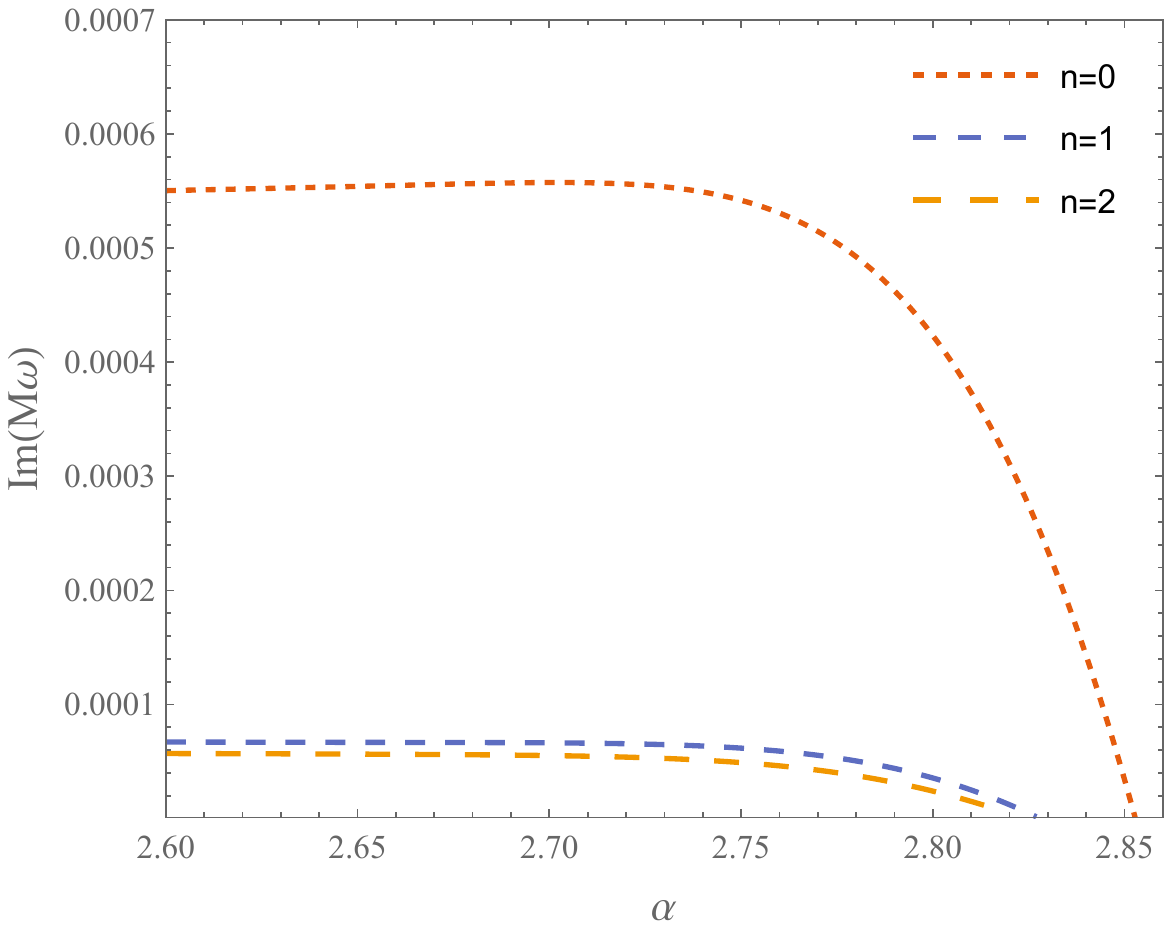}\quad
\includegraphics[width=0.46\textwidth]{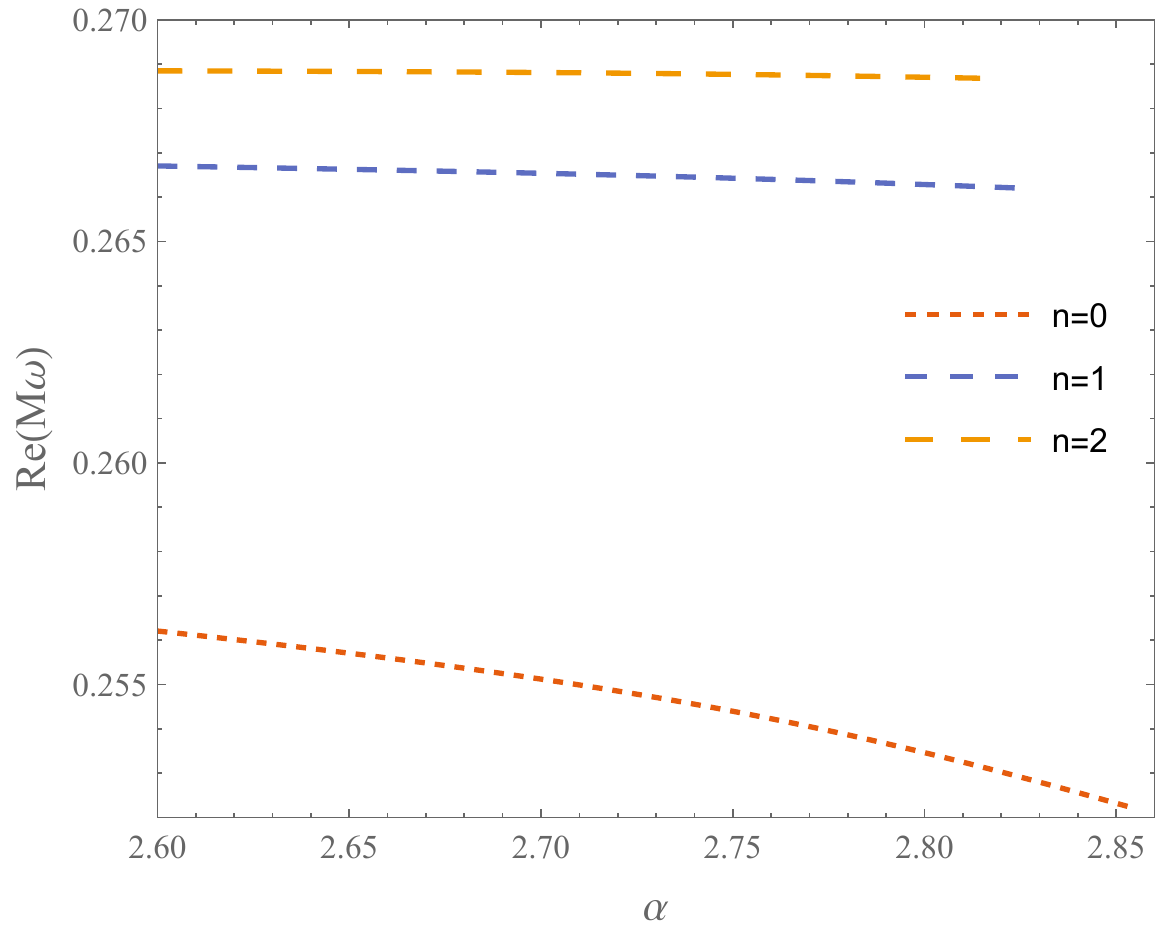}\\
\includegraphics[width=0.46\textwidth]{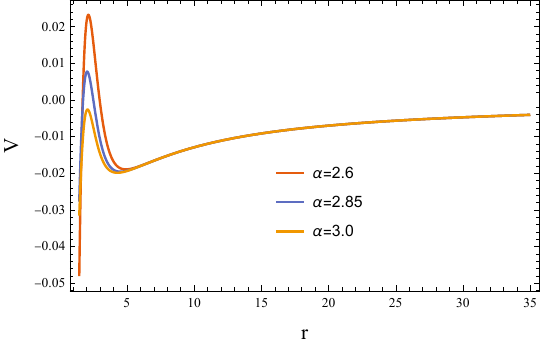}
\caption{{\em Top}: Spectrum of the superradiant unstable modes for
 $n=0,1$ and $2$, as a function of $\alpha$.  {\em Bottom}: Profile of the effective potential $V$ for various $\alpha$. Here we set $qM^2=1.8$ and $M\mu=0.27$.}
\label{a}
\end{figure}

\begin{figure}[!htbp]
\centering
\includegraphics[width=0.46\textwidth]{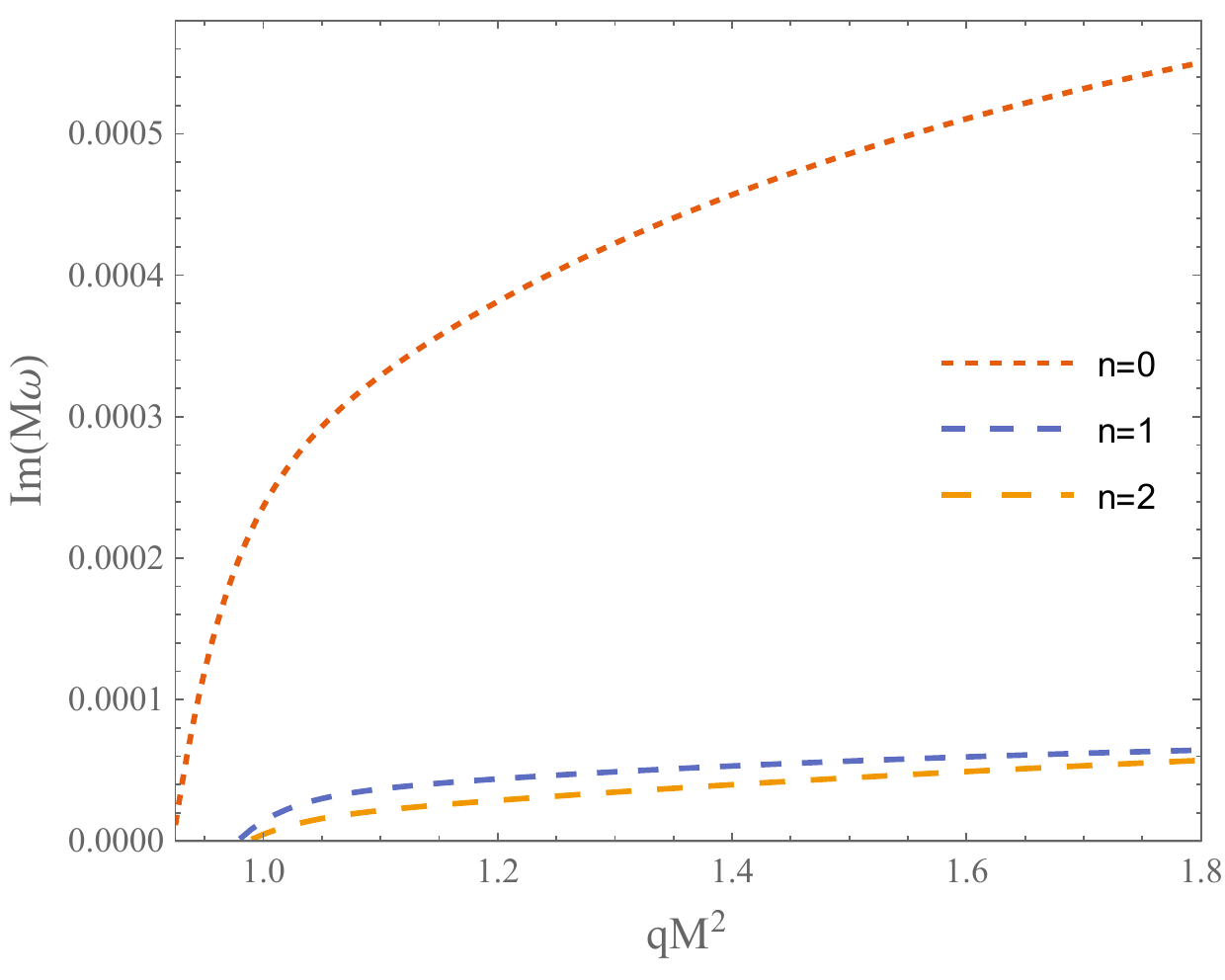}\quad
\includegraphics[width=0.46\textwidth]{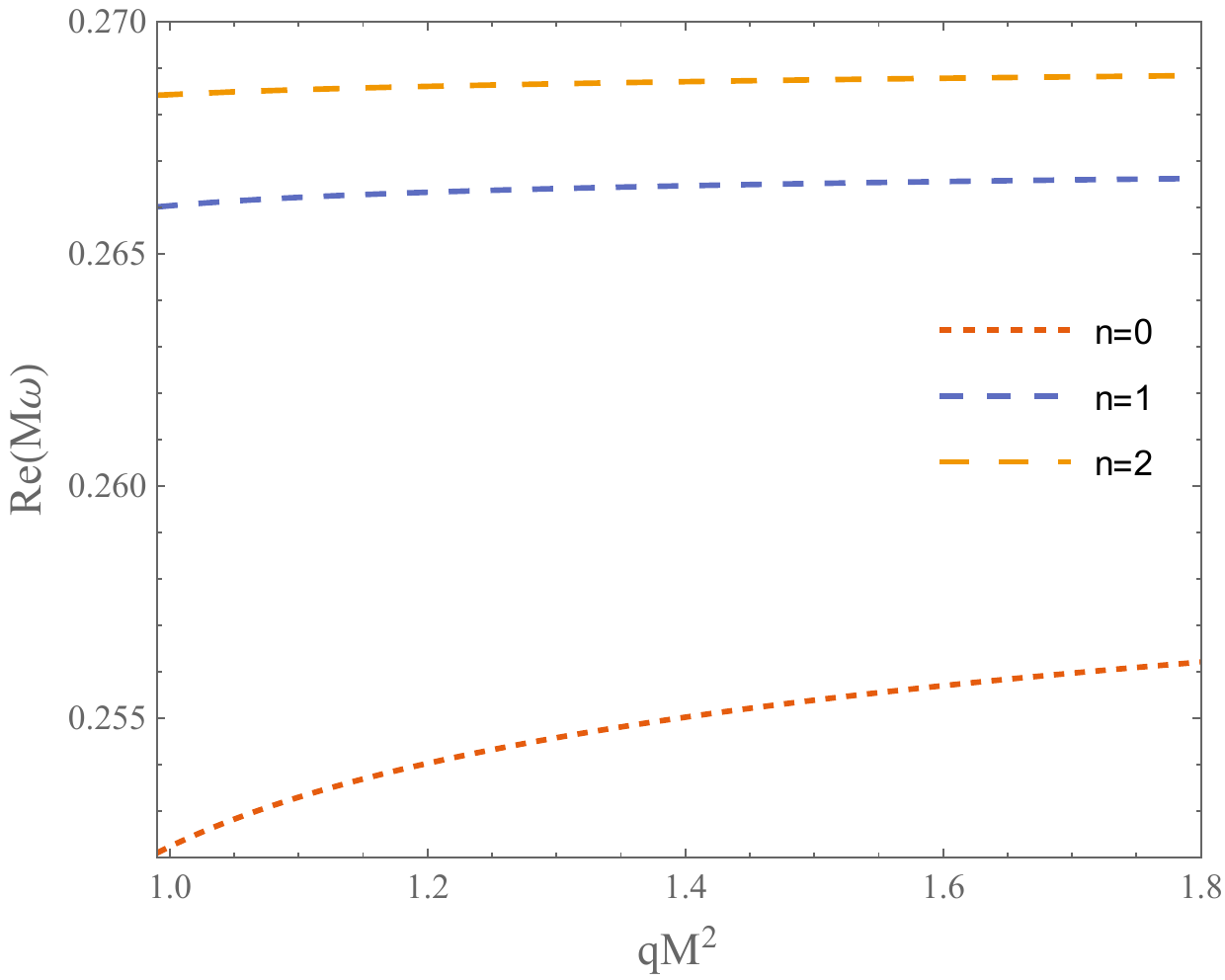}\\
\includegraphics[width=0.46\textwidth]{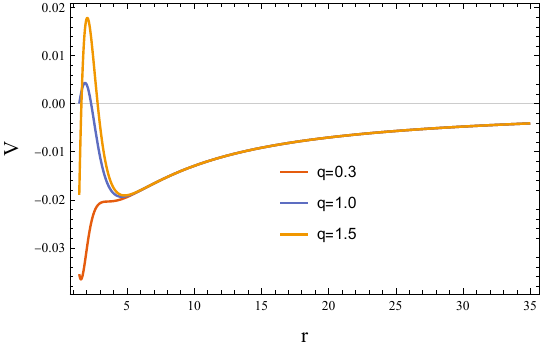}
\caption{{\em Top}: Spectrum of the superradiant unstable modes for
 $n=0,1$ and $2$, as a function of $q$. {\em Bottom}: Profile of the effective potential $V$ for various $q$. Here $\alpha=2.6$ and $M\mu=0.27$.}
\label{q}
\end{figure}

\begin{figure}[!htbp]
\centering
\includegraphics[width=0.46\textwidth]{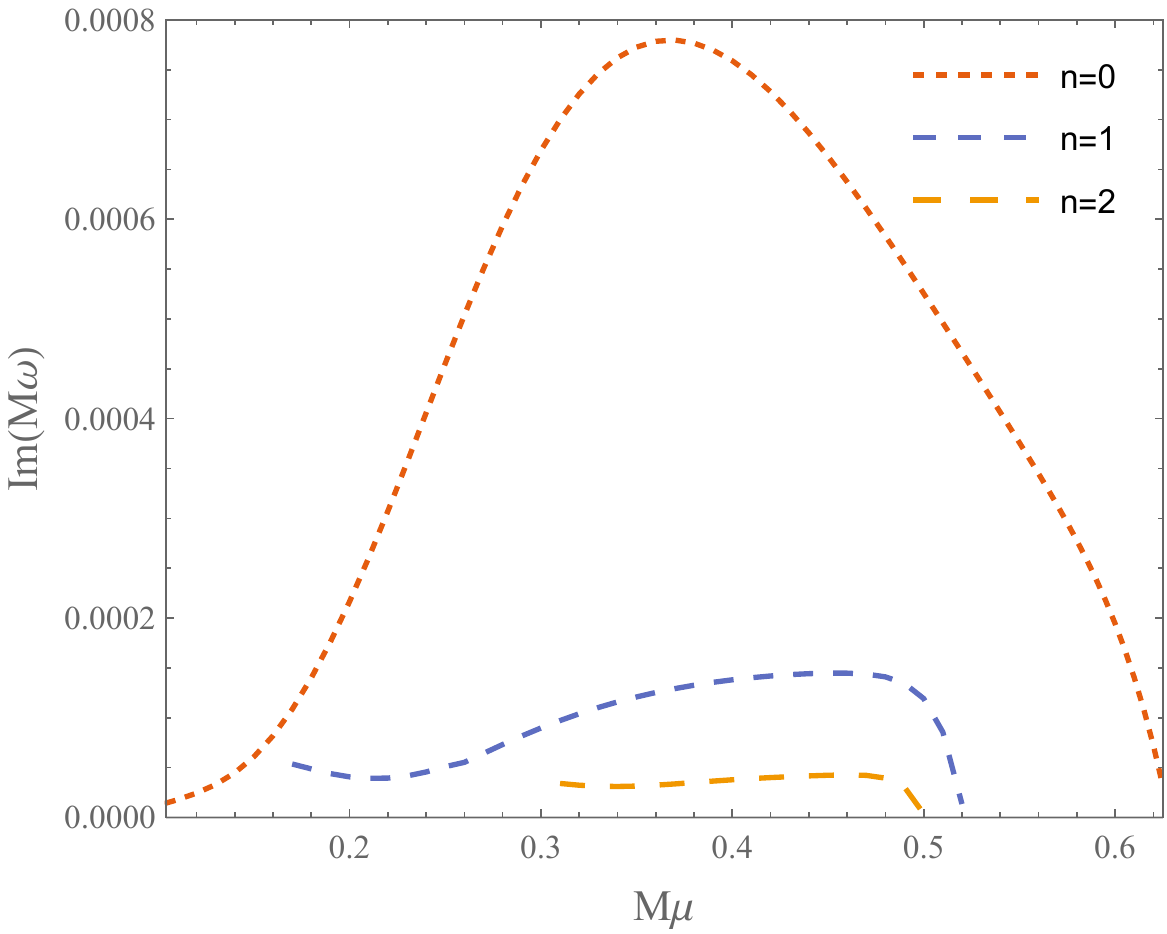}\quad
\includegraphics[width=0.46\textwidth]{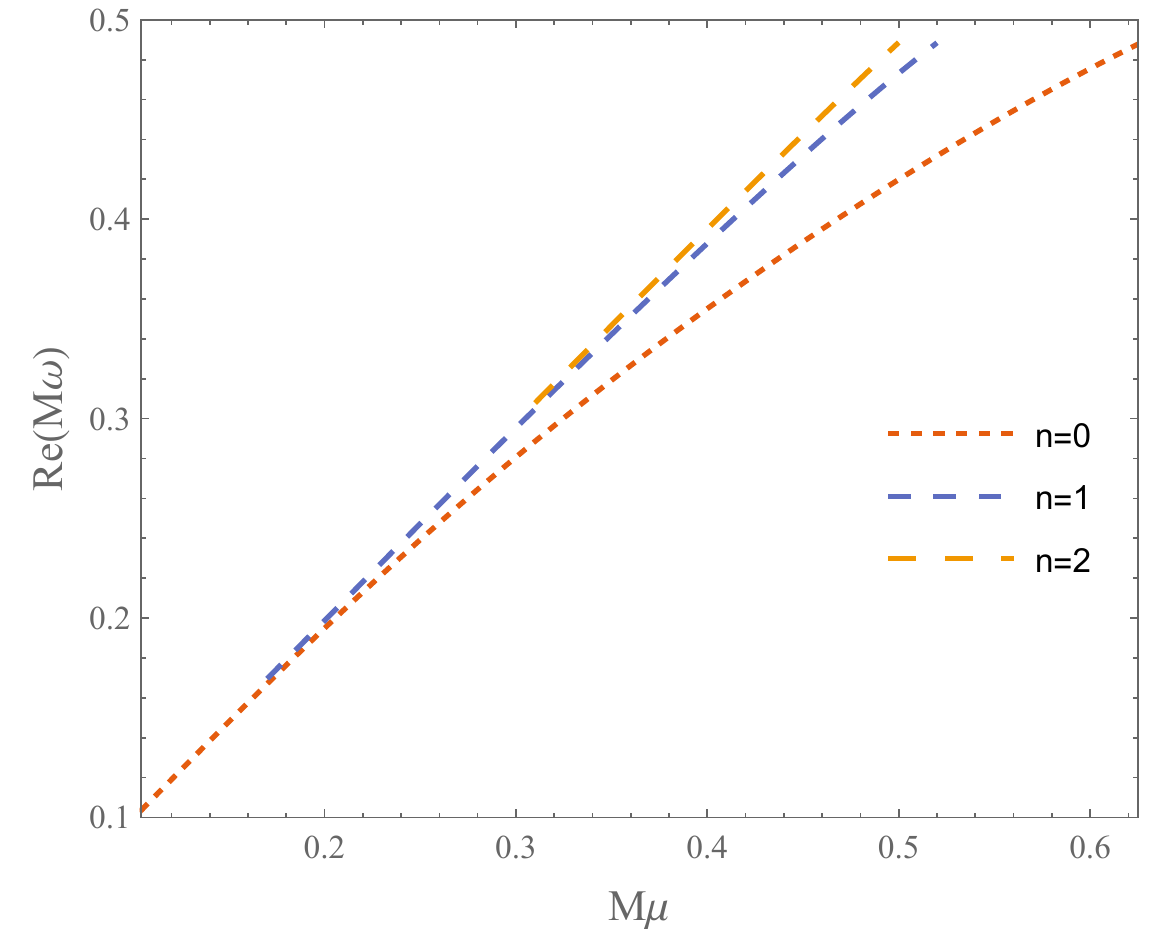}\\
\includegraphics[width=0.46\textwidth]{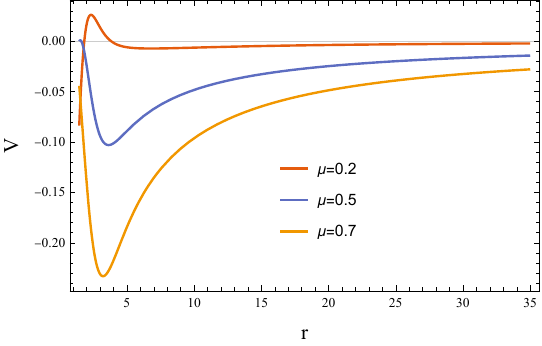}
\caption{{\em Top}: Spectrum of the superradiant unstable moeds for
$n=0,1$ and $2$, as a function of $\mu$. {\em Bottom}: Profile of the effective potential $V$ for various $\mu$. Here $\alpha=2.6$ and $qM^2=1.8$.}
\label{mu}
\end{figure}

In Figs. \ref{a}, \ref{q} and \ref{mu}, we show the influences of the three parameters, $\{\alpha, q, \mu\}$, on the frequency $\omega$ of the superradiant unstable modes respectively. 

From Fig. \ref{a}, one can see once again that, for fixed $q$ and $\mu$, only when the deformation parameter $\alpha$ takes value in a narrow interval $\alpha \in [\alpha_c, \alpha_0]$ can there exist superradiant instability, i.e., the imaginary part of the frequency $\text{Im} \omega>0$. The lower limit $\alpha_c \simeq 2.56$ as we have mentioned above, while the upper limit $\alpha_0$ depends on the values of other various parameters. This is expected as $\alpha$ becomes larger and larger, the spacetime becomes more and more like the standard Schwarzschild and it has been well-known that the Schwarzschild black hole does not exhibit superradiant instability. Moreover, as $\alpha$ increases in the interval $[\alpha_c, \alpha_0]$,  $\text{Im} \omega$ decreases monotonically and thus the instability becomes weaker. This can be understood from the effective potential $V$ shown in the bottom panel. The height of the potential barrier near the horizon decreases with the increase of $\alpha$, this makes it easier for the wave to fall into black holes, making it more difficult to form bound states. Moreover, as $\alpha$ exceeds the upper limit $\alpha_0$ (for example $\alpha = 3.0$ in the bottom panel), the height of the potential barrier is lower than zero and thus bound states can not be formed. The maximum growing rate of the instability for the fundamental mode $n=0$ is $\text{Im} (M \omega) \sim 5.8 \times 10^{-4}$. The real part of the frequency $\text{Re} \omega$ is very little affected by $\alpha$, especially for higher overtones. This can also be understood from the hydrogenic approximation, as the deformation of the spacetime geometry by $\alpha$ is mainly in the near-horizon region and modes of higher overtones live  far away from the horizon. 

From Fig. \ref{q}, one can see once again that only when the charge parameter of the scalar field $q$ exceeds some threshold value $q_c$ can there exist superradiant instability. Moreover, as $q$ increases,  $\text{Im} \omega$ increases monotonically which implies more violent instability. These can also be understood from the effective potential shown in the bottom panel. When $q<q_c$ (for example $q=0.3$ in the bottom panel), the height of the potential barrier is lower than zero, making the bound state hard to form. And as $q$ is increased, the height of the potential barrier becomes higher, making the bound state easier to form. $\text{Re} \omega$ is very little affected by $q$, especially for higher overtones. This can also be understood from the hydrogenic approximation. 

From Fig. \ref{mu}, one can see once again that only when the mass parameter $\mu$ takes value in a certain interval,  $\mu \in [\mu_c, \mu_0]$,  can there exist superradiant instability. The dependence of $\text{Im} \omega$ on the mass parameter of the scalar field $\mu$ is complicated. For the fundamental modes $n=0$, as $\mu$ increases, $\text{Im} \omega$ first increases and then decreases and takes a maximum value $\text{Im} (M \omega) \sim 7.8 \times 10^{-4}$ at $M \mu \sim 0.37$. For overtones (for example, $n=1$ and $2$), as $\mu$ increases, $\text{Im} \omega$ first decreases and then increases and then decreases, taking a maximum value at some certain $\mu$. These can be understood from the effective potential shown in the bottom panel. As $\mu$ is increased, the potential well becomes deeper which is favorable for the formation of bound states, while the potential barrier becomes lower which is unfavorable for the formation of bound states. Under the competition of the two factors, $\text{Im} \omega$ will take a maximum value at some mediate value. Rather differently, $\text{Re} \omega$ increases monotonically with the increase of $\mu$ for any $n$. This can also be understood from the hydrogenic approximation. 

The analysis depicted in Figs. \ref{a}, \ref{q}, and \ref{mu} reveals that the growth rate order can reach approximately $\text{Im} (M \omega) \sim 5 \times 10^{-4}$ for $n=0$ and $\text{Im} (M \omega) \sim 5 \times 10^{-5}$ for $n=1, 2$. These values surpass the observed growth rates in ABG black holes, where the maximum growth rate is approximately $\text{Im} (M \omega) \sim 8 \times 10^{-6}$ for $n=0$ as reported in \cite{Dolan:2024qqr}, and notably exceed the growth rates in Kerr black holes under massive scalar field perturbations, where the maximum growth rate is around $\text{Im} (M \omega) \sim 10^{-7}$ for $n=0$ \cite{Dolan:2007mj}.

\section{Summary and discussions}

The investigation presented here delves into the charged superradiant instability affecting a spherical regular black hole when subjected to scalar field perturbations. The spacetime model, as detailed in \cite{Ovalle:2023ref}, is formulated using the GD technique. This model smoothly transitions between Minkowski spacetime and the standard Schwarzschild black hole by introducing a parameter $\alpha$ that governs the deformation of the spacetime. Unlike other regular black holes such as the ABG black holes \cite{Ayon-Beato:1998hmi,Ayon-Beato:1999kuh}, the deformation of this black hole from the standard Schwarzschild form shows an exponential decay along the radial direction. Through a combination of semi-analytical and numerical approaches, it has been observed that this particular black hole can exhibit charged superradiant instability within specific parameter regimes. Recent studies in Refs. \cite{Hod:2024aen, dePaula:2024xnd, Dolan:2024qqr} have already highlighted the presence of charged superradiant instability in ABG regular black holes. The findings of our current research further support the notion that regular black holes may generally manifest superradiant instability.

The investigation considers the critical conditions necessary to initiate superradiant instability, providing an estimation of the permissible range for the presence of such unstable modes in parameter spaces, as depicted in Fig. \ref{AllowedRegion}. Subsequently, through the utilization of a numerical technique involving direct integration, the frequencies associated with the superradiant unstable modes are determined and presented in Figs. \ref{a}, \ref{q}, and \ref{mu}. The analysis of these figures reveals the impact of the parameters $\{\alpha, q, \mu\}$ on the onset and intensity of the instability. Notably, it is observed that: (i) Instability is only induced within a specific range of $\alpha$ values, $\alpha \in [\alpha_c, \alpha_0]$, with the growth rate $\text{Im} (M \omega)$ exhibiting a decreasing trend as $\alpha$ increases; (ii) superradiant instability manifests only when $q$ surpasses a critical threshold $q_c$, with the growth rate $\text{Im} (M \omega)$ demonstrating a consistent rise with increasing $q$; (iii) Instability arises within a designated interval of $\mu$ values, $\mu \in [\mu_c, \mu_0]$, with the growth rate $\text{Im} (M \omega)$ showcasing a complex dependence on $\mu$, reaching a peak at an intermediary $\mu$ value with order approximately $M \mu \sim 0.5$. Restoring physical units, one has \cite{Dolan:2024qqr}
\begin{equation}
    M \mu \equiv \frac{G M \mu}{\hbar c}\sim 7.5 \times 10^9 \left(\frac{M}{M_\odot}\right)\left(\frac{\mu}{1 \text{eV}}\right).
\end{equation}
This implies that the superradiant instability becomes notable when the Compton wavelength of a field is on par with the gravitational radius of a black hole. Although this instability is usually insignificant for Standard Model fields interacting with astrophysical black holes, it can manifest in ultralight fields such as axions \cite{Arvanitaki:2009fg,Arvanitaki:2010sy} or in scenarios where heavier fields coupled to primordial black holes \cite{Pani:2013hpa,Ferraz:2020zgi,Branco:2023frw,Calza:2023rjt}. See more discussions on its astrophysical implications, please refer to the review \cite{Brito:2015oca} and references therein. The qualitative explanation of the influence of these parameters on the growth rate can be elucidated by examining the characteristics of the effective potential $V(r)$.

The growth rate order, revealed in Figs. \ref{a}, \ref{q} and \ref{mu}, can reach approximately $\text{Im} (M \omega) \sim 5 \times 10^{-4}$ for $n=0$ and $\text{Im} (M \omega) \sim 5 \times 10^{-5}$ for $n=1, 2$. These values surpass the observed growth rates in ABG black holes, where the maximum growth rate is approximately $\text{Im} (M \omega) \sim 8 \times 10^{-6}$ for $n=0$ as reported in \cite{Dolan:2024qqr}, and notably exceed the growth rates in Kerr black holes under massive scalar field perturbations, where the maximum growth rate is around $\text{Im} (M \omega) \sim 10^{-7}$ for $n=0$ \cite{Dolan:2007mj}. Restoring physical units, we have
\begin{equation}
     \text{Im} (M \omega) \approx 4.9 \times 10^{-6} \left(\frac{M}{M_\odot}\right) \left(\frac{\text{Im} \omega}{1 \text{Hz}}\right).
\end{equation}
The instability time scale $\tau \equiv 1/ \text{Im} \omega$ will be notably short in astrophysical scenarios, rendering the phenomenon observable.

The findings suggest that superradiant instabilities could potentially be prevalent in regular black holes. Given that regular black holes are typically sustained by a nonlinear electromagnetic field as a source, there is a basis for considering that the superradiant instability may be intricately linked to the nonlinear aspects of the electromagnetic field. To confirm this hypothesis, a broader assessment of black holes (both regular and singular) with nonlinear electromagnetic fields as sources is necessary to investigate the presence of superradiant instability. Moreover, the presence of superradiant instability poses a great challenge to the stability of regular black holes. The final destination of regular black holes under such instabilities remains an open question, necessitating nonlinear numerical simulations to study the evolution of perturbations.

\begin{acknowledgments}

	This work is supported by the National Natural Science Foundation of China (NNSFC) under Grant No 12075207.

\end{acknowledgments}

\bibliographystyle{utphys}
\bibliography{ref}

\providecommand{\href}[2]{#2}\begingroup\raggedright\begin{thebibliography}{10}

\bibitem{LIGOScientific:2016aoc}
{\bfseries LIGO Scientific, Virgo} Collaboration, B.~P. Abbott {\em et~al.}, ``{Observation of Gravitational Waves from a Binary Black Hole Merger},'' \href{https://dx.doi.org/10.1103/PhysRevLett.116.061102}{{\em Phys. Rev. Lett.} {\bfseries 116} no.~6, (2016) 061102}, \href{https://arxiv.org/abs/1602.03837}{{\ttfamily arXiv:1602.03837 [gr-qc]}}.

\bibitem{LIGOScientific:2016sjg}
{\bfseries LIGO Scientific, Virgo} Collaboration, B.~P. Abbott {\em et~al.}, ``{GW151226: Observation of Gravitational Waves from a 22-Solar-Mass Binary Black Hole Coalescence},'' \href{https://dx.doi.org/10.1103/PhysRevLett.116.241103}{{\em Phys. Rev. Lett.} {\bfseries 116} no.~24, (2016) 241103}, \href{https://arxiv.org/abs/1606.04855}{{\ttfamily arXiv:1606.04855 [gr-qc]}}.

\bibitem{EventHorizonTelescope:2019dse}
{\bfseries Event Horizon Telescope} Collaboration, K.~Akiyama {\em et~al.}, ``{First M87 Event Horizon Telescope Results. I. The Shadow of the Supermassive Black Hole},'' \href{https://dx.doi.org/10.3847/2041-8213/ab0ec7}{{\em Astrophys. J. Lett.} {\bfseries 875} (2019) L1}, \href{https://arxiv.org/abs/1906.11238}{{\ttfamily arXiv:1906.11238 [astro-ph.GA]}}.

\bibitem{EventHorizonTelescope:2019ggy}
{\bfseries Event Horizon Telescope} Collaboration, K.~Akiyama {\em et~al.}, ``{First M87 Event Horizon Telescope Results. VI. The Shadow and Mass of the Central Black Hole},'' \href{https://dx.doi.org/10.3847/2041-8213/ab1141}{{\em Astrophys. J. Lett.} {\bfseries 875} no.~1, (2019) L6}, \href{https://arxiv.org/abs/1906.11243}{{\ttfamily arXiv:1906.11243 [astro-ph.GA]}}.

\bibitem{Tolman:1939jz}
R.~C. Tolman, ``{Static solutions of Einstein's field equations for spheres of fluid},'' \href{https://dx.doi.org/10.1103/PhysRev.55.364}{{\em Phys. Rev.} {\bfseries 55} (1939) 364--373}.

\bibitem{Oppenheimer:1939ue}
J.~R. Oppenheimer and H.~Snyder, ``{On Continued gravitational contraction},'' \href{https://dx.doi.org/10.1103/PhysRev.56.455}{{\em Phys. Rev.} {\bfseries 56} (1939) 455--459}.

\bibitem{Joshi:2011rlc}
P.~S. Joshi and D.~Malafarina, ``{Recent developments in gravitational collapse and spacetime singularities},'' \href{https://dx.doi.org/10.1142/S0218271811020792}{{\em Int. J. Mod. Phys. D} {\bfseries 20} (2011) 2641--2729}, \href{https://arxiv.org/abs/1201.3660}{{\ttfamily arXiv:1201.3660 [gr-qc]}}.

\bibitem{Christodoulou:1984mz}
D.~Christodoulou, ``{Violation of cosmic censorship in the gravitational collapse of a dust cloud},'' \href{https://dx.doi.org/10.1007/BF01223743}{{\em Commun. Math. Phys.} {\bfseries 93} (1984) 171--195}.

\bibitem{Penrose:1969pc}
R.~Penrose, ``{Gravitational collapse: The role of general relativity},'' \href{https://dx.doi.org/10.1023/A:1016578408204}{{\em Riv. Nuovo Cim.} {\bfseries 1} (1969) 252--276}.

\bibitem{Joshi:2008zz}
P.~S. Joshi, ed., \href{https://dx.doi.org/10.1017/CBO9780511536274}{{\em {Gravitational Collapse and Spacetime Singularities}}}.
\newblock Cambridge Monographs on Mathematical Physics. Cambridge University Press, 9, 2012.

\bibitem{Penrose:1964wq}
R.~Penrose, ``{Gravitational collapse and space-time singularities},'' \href{https://dx.doi.org/10.1103/PhysRevLett.14.57}{{\em Phys. Rev. Lett.} {\bfseries 14} (1965) 57--59}.

\bibitem{Hawking:1973uf}
S.~W. Hawking and G.~F.~R. Ellis, \href{https://dx.doi.org/10.1017/9781009253161}{{\em {The Large Scale Structure of Space-Time}}}.
\newblock Cambridge Monographs on Mathematical Physics. Cambridge University Press, 2, 2023.

\bibitem{Senovilla:1998oua}
J.~M.~M. Senovilla, ``{Singularity Theorems and Their Consequences},'' \href{https://dx.doi.org/10.1023/A:1018801101244}{{\em Gen. Rel. Grav.} {\bfseries 30} (1998) 701}, \href{https://arxiv.org/abs/1801.04912}{{\ttfamily arXiv:1801.04912 [gr-qc]}}.

\bibitem{1968qtr..conf...87B}
J.~{Bardeen}, ``{Non-singular general relativistic gravitational collapse},'' in {\em Proceedings of the 5th International Conference on Gravitation and the Theory of Relativity}, p.~87.
\newblock Sept., 1968.

\bibitem{Ayon-Beato:1998hmi}
E.~{Ay{\'o}n-Beato} and A.~Garc{\'i}a, ``Regular black hole in general relativity coupled to nonlinear electrodynamics,'' \href{https://dx.doi.org/10.1103/PhysRevLett.80.5056}{{\em Phys. Rev. Lett.} {\bfseries 80} no.~23, (1998) 5056--5059}, \href{https://arxiv.org/abs/gr-qc/9911046}{{\ttfamily arXiv:gr-qc/9911046}}.

\bibitem{Ayon-Beato:1999kuh}
E.~{Ay{\'o}n-Beato} and A.~Garc{\'{\i}}a, ``New regular black hole solution from nonlinear electrodynamics,'' \href{https://dx.doi.org/10.1016/S0370-2693(99)01038-2}{{\em Physics Letters B} {\bfseries 464} no.~1-2, (1999) 25}, \href{https://arxiv.org/abs/hep-th/9911174}{{\ttfamily arXiv:hep-th/9911174}}.

\bibitem{Ayon-Beato:2000mjt}
E.~{Ay{\'o}n-Beato} and A.~Garc{\'{\i}}a, ``The bardeen model as a nonlinear magnetic monopole,'' \href{https://dx.doi.org/10.1016/S0370-2693(00)01125-4}{{\em Physics Letters B} {\bfseries 493} no.~1-2, (2000) 149--152}, \href{https://arxiv.org/abs/gr-qc/0009077}{{\ttfamily arXiv:gr-qc/0009077}}.

\bibitem{Fan:2016hvf}
Z.-Y. Fan and X.~Wang, ``{Construction of Regular Black Holes in General Relativity},'' \href{https://dx.doi.org/10.1103/PhysRevD.94.124027}{{\em Phys. Rev. D} {\bfseries 94} no.~12, (2016) 124027}, \href{https://arxiv.org/abs/1610.02636}{{\ttfamily arXiv:1610.02636 [gr-qc]}}.

\bibitem{Ghosh:2022gka}
R.~Ghosh, M.~Rahman, and A.~K. Mishra, ``{Regularized stable Kerr black hole: cosmic censorships, shadow and quasi-normal modes},'' \href{https://dx.doi.org/10.1140/epjc/s10052-023-11252-0}{{\em Eur. Phys. J. C} {\bfseries 83} no.~1, (2023) 91}, \href{https://arxiv.org/abs/2209.12291}{{\ttfamily arXiv:2209.12291 [gr-qc]}}.

\bibitem{Lan:2023cvz}
C.~Lan, H.~Yang, Y.~Guo, and Y.-G. Miao, ``{Regular Black Holes: A Short Topic Review},'' \href{https://dx.doi.org/10.1007/s10773-023-05454-1}{{\em Int. J. Theor. Phys.} {\bfseries 62} no.~9, (2023) 202}, \href{https://arxiv.org/abs/2303.11696}{{\ttfamily arXiv:2303.11696 [gr-qc]}}.

\bibitem{Torres:2022twv}
R.~Torres, ``{Regular Rotating Black Holes: A Review},'' \href{https://arxiv.org/abs/2208.12713}{{\ttfamily arXiv:2208.12713 [gr-qc]}}.

\bibitem{Bronnikov:2022ofk}
K.~A. Bronnikov, ``{Regular black holes sourced by nonlinear electrodynamics},'' \href{https://arxiv.org/abs/2211.00743}{{\ttfamily arXiv:2211.00743 [gr-qc]}}.

\bibitem{Bueno:2024dgm}
P.~Bueno, P.~A. Cano, and R.~A. Hennigar, ``{Regular Black Holes From Pure Gravity},'' \href{https://arxiv.org/abs/2403.04827}{{\ttfamily arXiv:2403.04827 [gr-qc]}}.

\bibitem{Heisenberg:1936nmg}
W.~Heisenberg and H.~Euler, ``{Consequences of Dirac's theory of positrons},'' \href{https://dx.doi.org/10.1007/BF01343663}{{\em Z. Phys.} {\bfseries 98} no.~11-12, (1936) 714--732}, \href{https://arxiv.org/abs/physics/0605038}{{\ttfamily arXiv:physics/0605038}}.

\bibitem{Born:1934gh}
M.~Born and L.~Infeld, ``{Foundations of the new field theory},'' \href{https://dx.doi.org/10.1098/rspa.1934.0059}{{\em Proc. Roy. Soc. Lond. A} {\bfseries 144} no.~852, (1934) 425--451}.

\bibitem{Polchinski:1998rq}
J.~Polchinski, \href{https://dx.doi.org/10.1017/CBO9780511816079}{{\em {String theory. Vol. 1: An introduction to the bosonic string}}}.
\newblock Cambridge Monographs on Mathematical Physics. Cambridge University Press, 12, 2007.

\bibitem{Polchinski:1998rr}
J.~Polchinski, \href{https://dx.doi.org/10.1017/CBO9780511618123}{{\em {String theory. Vol. 2: Superstring theory and beyond}}}.
\newblock Cambridge Monographs on Mathematical Physics. Cambridge University Press, 12, 2007.

\bibitem{Allahyari:2019jqz}
A.~Allahyari, M.~Khodadi, S.~Vagnozzi, and D.~F. Mota, ``{Magnetically charged black holes from non-linear electrodynamics and the Event Horizon Telescope},'' \href{https://dx.doi.org/10.1088/1475-7516/2020/02/003}{{\em JCAP} {\bfseries 02} (2020) 003}, \href{https://arxiv.org/abs/1912.08231}{{\ttfamily arXiv:1912.08231 [gr-qc]}}.

\bibitem{Vagnozzi:2022moj}
S.~Vagnozzi {\em et~al.}, ``{Horizon-scale tests of gravity theories and fundamental physics from the Event Horizon Telescope image of Sagittarius A},'' \href{https://dx.doi.org/10.1088/1361-6382/acd97b}{{\em Class. Quant. Grav.} {\bfseries 40} no.~16, (2023) 165007}, \href{https://arxiv.org/abs/2205.07787}{{\ttfamily arXiv:2205.07787 [gr-qc]}}.

\bibitem{Cai:2020kue}
X.-C. Cai and Y.-G. Miao, ``{Quasinormal modes of the generalized Ay\'on-Beato\textendash{}Garc\'\i{}a black hole in scalar-tensor-vector gravity},'' \href{https://dx.doi.org/10.1103/PhysRevD.102.084061}{{\em Phys. Rev. D} {\bfseries 102} no.~8, (2020) 084061}, \href{https://arxiv.org/abs/2008.04576}{{\ttfamily arXiv:2008.04576 [gr-qc]}}.

\bibitem{Cai:2021ele}
X.-C. Cai and Y.-G. Miao, ``{Quasinormal modes and shadows of a new family of Ay\'on-Beato-Garc\'\i{}a black holes},'' \href{https://dx.doi.org/10.1103/PhysRevD.103.124050}{{\em Phys. Rev. D} {\bfseries 103} no.~12, (2021) 124050}, \href{https://arxiv.org/abs/2104.09725}{{\ttfamily arXiv:2104.09725 [gr-qc]}}.

\bibitem{Yang:2022uze}
H.~Yang and Y.-G. Miao, ``{Superradiance of massive scalar particles around rotating regular black holes*},'' \href{https://dx.doi.org/10.1088/1674-1137/accdc7}{{\em Chin. Phys. C} {\bfseries 47} no.~7, (2023) 075101}, \href{https://arxiv.org/abs/2211.15130}{{\ttfamily arXiv:2211.15130 [gr-qc]}}.

\bibitem{Meng:2022oxg}
K.~Meng and S.-J. Zhang, ``{Gravito-electromagnetic perturbations and QNMs of regular black holes},'' \href{https://dx.doi.org/10.1088/1361-6382/acf3c6}{{\em Class. Quant. Grav.} {\bfseries 40} no.~19, (2023) 195024}, \href{https://arxiv.org/abs/2210.00295}{{\ttfamily arXiv:2210.00295 [gr-qc]}}.

\bibitem{Pedrotti:2024znu}
D.~Pedrotti and S.~Vagnozzi, ``{Quasinormal modes-shadow correspondence for rotating regular black holes},'' \href{https://dx.doi.org/10.1103/PhysRevD.110.084075}{{\em Phys. Rev. D} {\bfseries 110} no.~8, (2024) 084075}, \href{https://arxiv.org/abs/2404.07589}{{\ttfamily arXiv:2404.07589 [gr-qc]}}.

\bibitem{Hod:2024aen}
S.~Hod, ``Spatially regular charged black holes supporting charged massive scalar clouds,'' \href{https://dx.doi.org/10.1103/PhysRevD.109.064074}{{\em Phys. Rev. D} {\bfseries 109} no.~6, (Mar., 2024) 064074}, \href{https://arxiv.org/abs/2401.07907}{{\ttfamily arXiv:2401.07907 [gr-qc]}}.

\bibitem{dePaula:2024xnd}
M.~A.~A. {de Paula}, L.~C.~S. Leite, S.~R. Dolan, and L.~C.~B. Crispino, ``Absorption and (unbounded) superradiance in a static regular black hole spacetime,'' \href{https://arxiv.org/abs/2401.01767}{{\ttfamily arXiv:2401.01767 [gr-qc]}}.

\bibitem{Dolan:2024qqr}
S.~R. Dolan, M.~A.~A. {de Paula}, L.~C.~S. Leite, and L.~C.~B. Crispino, ``Superradiant instability of a charged regular black hole,'' \href{https://dx.doi.org/10.1103/PhysRevD.109.124037}{{\em Phys. Rev. D} {\bfseries 109} no.~12, (June, 2024) 124037}, \href{https://arxiv.org/abs/2401.14967}{{\ttfamily arXiv:2401.14967 [gr-qc]}}.

\bibitem{Damour:1976kh}
T.~Damour, N.~Deruelle, and R.~Ruffini, ``{On Quantum Resonances in Stationary Geometries},'' \href{https://dx.doi.org/10.1007/BF02725534}{{\em Lett. Nuovo Cim.} {\bfseries 15} (1976) 257--262}.

\bibitem{Detweiler:1980uk}
S.~L. Detweiler, ``{KLEIN-GORDON EQUATION AND ROTATING BLACK HOLES},'' \href{https://dx.doi.org/10.1103/PhysRevD.22.2323}{{\em Phys. Rev. D} {\bfseries 22} (1980) 2323--2326}.

\bibitem{Zouros:1979iw}
T.~J.~M. Zouros and D.~M. Eardley, ``{INSTABILITIES OF MASSIVE SCALAR PERTURBATIONS OF A ROTATING BLACK HOLE},'' \href{https://dx.doi.org/10.1016/0003-4916(79)90237-9}{{\em Annals Phys.} {\bfseries 118} (1979) 139--155}.

\bibitem{Cardoso:2005vk}
V.~Cardoso and S.~Yoshida, ``{Superradiant instabilities of rotating black branes and strings},'' \href{https://dx.doi.org/10.1088/1126-6708/2005/07/009}{{\em JHEP} {\bfseries 07} (2005) 009}, \href{https://arxiv.org/abs/hep-th/0502206}{{\ttfamily arXiv:hep-th/0502206}}.

\bibitem{Dolan:2007mj}
S.~R. Dolan, ``{Instability of the massive Klein-Gordon field on the Kerr spacetime},'' \href{https://dx.doi.org/10.1103/PhysRevD.76.084001}{{\em Phys. Rev. D} {\bfseries 76} (2007) 084001}, \href{https://arxiv.org/abs/0705.2880}{{\ttfamily arXiv:0705.2880 [gr-qc]}}.

\bibitem{Konoplya:2013rxa}
R.~A. Konoplya and A.~Zhidenko, ``{Massive charged scalar field in the Kerr-Newman background I: quasinormal modes, late-time tails and stability},'' \href{https://dx.doi.org/10.1103/PhysRevD.88.024054}{{\em Phys. Rev. D} {\bfseries 88} (2013) 024054}, \href{https://arxiv.org/abs/1307.1812}{{\ttfamily arXiv:1307.1812 [gr-qc]}}.

\bibitem{Hod:2012wmy}
S.~Hod, ``{Stability of the extremal Reissner-Nordstroem black hole to charged scalar perturbations},'' \href{https://dx.doi.org/10.1016/j.physletb.2012.06.043}{{\em Phys. Lett. B} {\bfseries 713} (2012) 505--508}, \href{https://arxiv.org/abs/1304.6474}{{\ttfamily arXiv:1304.6474 [gr-qc]}}.

\bibitem{Hod:2013nn}
S.~Hod, ``{No-bomb theorem for charged Reissner-Nordstroem black holes},'' \href{https://dx.doi.org/10.1016/j.physletb.2012.12.013}{{\em Phys. Lett. B} {\bfseries 718} (2013) 1489--1492}.

\bibitem{Herdeiro:2013pia}
C.~A.~R. Herdeiro, J.~C. Degollado, and H.~F. R\'unarsson, ``{Rapid growth of superradiant instabilities for charged black holes in a cavity},'' \href{https://dx.doi.org/10.1103/PhysRevD.88.063003}{{\em Phys. Rev. D} {\bfseries 88} (2013) 063003}, \href{https://arxiv.org/abs/1305.5513}{{\ttfamily arXiv:1305.5513 [gr-qc]}}.

\bibitem{Zhu:2014sya}
Z.~Zhu, S.-J. Zhang, C.~E. Pellicer, B.~Wang, and E.~Abdalla, ``{Stability of Reissner-Nordstr\"om black hole in de Sitter background under charged scalar perturbation},'' \href{https://dx.doi.org/10.1103/PhysRevD.90.044042}{{\em Phys. Rev. D} {\bfseries 90} no.~4, (2014) 044042}, \href{https://arxiv.org/abs/1405.4931}{{\ttfamily arXiv:1405.4931 [hep-th]}}. [Addendum: Phys.Rev.D 90, 049904 (2014)].

\bibitem{Konoplya:2014lha}
R.~A. Konoplya and A.~Zhidenko, ``{Charged scalar field instability between the event and cosmological horizons},'' \href{https://dx.doi.org/10.1103/PhysRevD.90.064048}{{\em Phys. Rev. D} {\bfseries 90} no.~6, (2014) 064048}, \href{https://arxiv.org/abs/1406.0019}{{\ttfamily arXiv:1406.0019 [hep-th]}}.

\bibitem{Dolan:2015dha}
S.~R. Dolan, S.~Ponglertsakul, and E.~Winstanley, ``{Stability of black holes in Einstein-charged scalar field theory in a cavity},'' \href{https://dx.doi.org/10.1103/PhysRevD.92.124047}{{\em Phys. Rev. D} {\bfseries 92} no.~12, (2015) 124047}, \href{https://arxiv.org/abs/1507.02156}{{\ttfamily arXiv:1507.02156 [gr-qc]}}.

\bibitem{Dias:2018zjg}
O.~J.~C. Dias and R.~Masachs, ``{Charged black hole bombs in a Minkowski cavity},'' \href{https://dx.doi.org/10.1088/1361-6382/aad70b}{{\em Class. Quant. Grav.} {\bfseries 35} no.~18, (2018) 184001}, \href{https://arxiv.org/abs/1801.10176}{{\ttfamily arXiv:1801.10176 [gr-qc]}}.

\bibitem{Davey:2021oye}
A.~Davey, O.~J.~C. Dias, and P.~Rodgers, ``{Phase diagram of the charged black hole bomb system},'' \href{https://dx.doi.org/10.1007/JHEP05(2021)189}{{\em JHEP} {\bfseries 05} (2021) 189}, \href{https://arxiv.org/abs/2103.12752}{{\ttfamily arXiv:2103.12752 [gr-qc]}}.

\bibitem{Richarte:2021fbi}
M.~G. Richarte, E.~L. Martins, and J.~C. Fabris, ``{Scattering and absorption of a scalar field impinging on a charged black hole in the Einstein-Maxwell-dilaton theory},'' \href{https://dx.doi.org/10.1103/PhysRevD.105.064043}{{\em Phys. Rev. D} {\bfseries 105} no.~6, (2022) 064043}, \href{https://arxiv.org/abs/2111.01595}{{\ttfamily arXiv:2111.01595 [gr-qc]}}.

\bibitem{Feiteira:2024awb}
D.~Feiteira, J.~P.~S. Lemos, and O.~B. Zaslavskii, ``{Penrose process in Reissner-Nordstr\"om-AdS black hole spacetimes: Black hole energy factories and black hole bombs},'' \href{https://dx.doi.org/10.1103/PhysRevD.109.064065}{{\em Phys. Rev. D} {\bfseries 109} no.~6, (2024) 064065}, \href{https://arxiv.org/abs/2401.13039}{{\ttfamily arXiv:2401.13039 [gr-qc]}}.

\bibitem{Arvanitaki:2009fg}
A.~Arvanitaki, S.~Dimopoulos, S.~Dubovsky, N.~Kaloper, and J.~March-Russell, ``{String Axiverse},'' \href{https://dx.doi.org/10.1103/PhysRevD.81.123530}{{\em Phys. Rev. D} {\bfseries 81} (2010) 123530}, \href{https://arxiv.org/abs/0905.4720}{{\ttfamily arXiv:0905.4720 [hep-th]}}.

\bibitem{Arvanitaki:2010sy}
A.~Arvanitaki and S.~Dubovsky, ``{Exploring the String Axiverse with Precision Black Hole Physics},'' \href{https://dx.doi.org/10.1103/PhysRevD.83.044026}{{\em Phys. Rev. D} {\bfseries 83} (2011) 044026}, \href{https://arxiv.org/abs/1004.3558}{{\ttfamily arXiv:1004.3558 [hep-th]}}.

\bibitem{Brito:2014wla}
R.~Brito, V.~Cardoso, and P.~Pani, ``{Black holes as particle detectors: evolution of superradiant instabilities},'' \href{https://dx.doi.org/10.1088/0264-9381/32/13/134001}{{\em Class. Quant. Grav.} {\bfseries 32} no.~13, (2015) 134001}, \href{https://arxiv.org/abs/1411.0686}{{\ttfamily arXiv:1411.0686 [gr-qc]}}.

\bibitem{Cunha:2015yba}
P.~V.~P. Cunha, C.~A.~R. Herdeiro, E.~Radu, and H.~F. Runarsson, ``{Shadows of Kerr black holes with scalar hair},'' \href{https://dx.doi.org/10.1103/PhysRevLett.115.211102}{{\em Phys. Rev. Lett.} {\bfseries 115} no.~21, (2015) 211102}, \href{https://arxiv.org/abs/1509.00021}{{\ttfamily arXiv:1509.00021 [gr-qc]}}.

\bibitem{Vincent:2016sjq}
F.~H. Vincent, E.~Gourgoulhon, C.~Herdeiro, and E.~Radu, ``{Astrophysical imaging of Kerr black holes with scalar hair},'' \href{https://dx.doi.org/10.1103/PhysRevD.94.084045}{{\em Phys. Rev. D} {\bfseries 94} no.~8, (2016) 084045}, \href{https://arxiv.org/abs/1606.04246}{{\ttfamily arXiv:1606.04246 [gr-qc]}}.

\bibitem{Cunha:2019ikd}
P.~V.~P. Cunha, C.~A.~R. Herdeiro, and E.~Radu, ``{EHT constraint on the ultralight scalar hair of the M87 supermassive black hole},'' \href{https://dx.doi.org/10.3390/universe5120220}{{\em Universe} {\bfseries 5} no.~12, (2019) 220}, \href{https://arxiv.org/abs/1909.08039}{{\ttfamily arXiv:1909.08039 [gr-qc]}}.

\bibitem{Creci:2020mfg}
G.~Creci, S.~Vandoren, and H.~Witek, ``{Evolution of black hole shadows from superradiance},'' \href{https://dx.doi.org/10.1103/PhysRevD.101.124051}{{\em Phys. Rev. D} {\bfseries 101} no.~12, (2020) 124051}, \href{https://arxiv.org/abs/2004.05178}{{\ttfamily arXiv:2004.05178 [gr-qc]}}.

\bibitem{Brito:2015oca}
R.~Brito, V.~Cardoso, and P.~Pani, ``{Superradiance}: {New Frontiers in Black Hole Physics},'' \href{https://dx.doi.org/10.1007/978-3-319-19000-6}{{\em Lect. Notes Phys.} {\bfseries 906} (2015) pp.1--237}, \href{https://arxiv.org/abs/1501.06570}{{\ttfamily arXiv:1501.06570 [gr-qc]}}.

\bibitem{Konoplya:2011qq}
R.~A. Konoplya and A.~Zhidenko, ``{Quasinormal modes of black holes: From astrophysics to string theory},'' \href{https://dx.doi.org/10.1103/RevModPhys.83.793}{{\em Rev. Mod. Phys.} {\bfseries 83} (2011) 793--836}, \href{https://arxiv.org/abs/1102.4014}{{\ttfamily arXiv:1102.4014 [gr-qc]}}.

\bibitem{Ovalle:2023ref}
J.~Ovalle, R.~Casadio, and A.~Giusti, ``{Regular hairy black holes through Minkowski deformation},'' \href{https://dx.doi.org/10.1016/j.physletb.2023.138085}{{\em Phys. Lett. B} {\bfseries 844} (2023) 138085}, \href{https://arxiv.org/abs/2304.03263}{{\ttfamily arXiv:2304.03263 [gr-qc]}}.

\bibitem{Ovalle:2017fgl}
J.~Ovalle, ``{Decoupling gravitational sources in general relativity: from perfect to anisotropic fluids},'' \href{https://dx.doi.org/10.1103/PhysRevD.95.104019}{{\em Phys. Rev. D} {\bfseries 95} no.~10, (2017) 104019}, \href{https://arxiv.org/abs/1704.05899}{{\ttfamily arXiv:1704.05899 [gr-qc]}}.

\bibitem{Ovalle:2018gic}
J.~Ovalle, ``{Decoupling gravitational sources in general relativity: The extended case},'' \href{https://dx.doi.org/10.1016/j.physletb.2018.11.029}{{\em Phys. Lett. B} {\bfseries 788} (2019) 213--218}, \href{https://arxiv.org/abs/1812.03000}{{\ttfamily arXiv:1812.03000 [gr-qc]}}.

\bibitem{Bekenstein:1973mi}
J.~D. Bekenstein, ``{Extraction of energy and charge from a black hole},'' \href{https://dx.doi.org/10.1103/PhysRevD.7.949}{{\em Phys. Rev. D} {\bfseries 7} (1973) 949--953}.

\bibitem{Pani:2013pma}
P.~Pani, ``{Advanced Methods in Black-Hole Perturbation Theory},'' \href{https://dx.doi.org/10.1142/S0217751X13400186}{{\em Int. J. Mod. Phys. A} {\bfseries 28} (2013) 1340018}, \href{https://arxiv.org/abs/1305.6759}{{\ttfamily arXiv:1305.6759 [gr-qc]}}.

\bibitem{Pani:2013hpa}
P.~Pani and A.~Loeb, ``{Constraining Primordial Black-Hole Bombs through Spectral Distortions of the Cosmic Microwave Background},'' \href{https://dx.doi.org/10.1103/PhysRevD.88.041301}{{\em Phys. Rev. D} {\bfseries 88} (2013) 041301}, \href{https://arxiv.org/abs/1307.5176}{{\ttfamily arXiv:1307.5176 [astro-ph.CO]}}.

\bibitem{Ferraz:2020zgi}
P.~B. Ferraz, T.~W. Kephart, and J.~a.~G. Rosa, ``{Superradiant pion clouds around primordial black holes},'' \href{https://dx.doi.org/10.1088/1475-7516/2022/07/026}{{\em JCAP} {\bfseries 07} no.~07, (2022) 026}, \href{https://arxiv.org/abs/2004.11303}{{\ttfamily arXiv:2004.11303 [gr-qc]}}.

\bibitem{Branco:2023frw}
N.~P. Branco, R.~Z. Ferreira, and J.~a.~G. Rosa, ``{Superradiant axion clouds around asteroid-mass primordial black holes},'' \href{https://dx.doi.org/10.1088/1475-7516/2023/04/003}{{\em JCAP} {\bfseries 04} (2023) 003}, \href{https://arxiv.org/abs/2301.01780}{{\ttfamily arXiv:2301.01780 [hep-ph]}}.

\bibitem{Calza:2023rjt}
M.~Calz\`a, J.~a.~G. Rosa, and F.~Serrano, ``{Primordial black hole superradiance and evaporation in the string axiverse},'' \href{https://dx.doi.org/10.1007/JHEP05(2024)140}{{\em JHEP} {\bfseries 05} (2024) 140}, \href{https://arxiv.org/abs/2306.09430}{{\ttfamily arXiv:2306.09430 [hep-ph]}}.

\end{thebibliography}\endgroup

\end{document}